\documentclass[12pt]{article}
\usepackage{amsmath,amssymb,graphicx,epsfig}

\def \lsim{\mathrel{\vcenter
{\hbox{$<$}\nointerlineskip\hbox{$\sim$}}}}
\def \gsim{\mathrel{\vcenter
{\hbox{$>$}\nointerlineskip\hbox{$\sim$}}}}

\def\bea{\begin{eqnarray}}
\def\eea{\end{eqnarray}}
\def\be{\begin{equation}}
\def\ee{\end{equation}}
\def\ba{\begin{array}}
\def\ea{\end{array}}
\def\nn{\nonumber}

\font\tenrsfs=rsfs10
\font\sevenrsfs=rsfs7
\font\fiversfs=rsfs5
\newfam\rsfsfam
\textfont\rsfsfam=\tenrsfs
\scriptfont\rsfsfam=\sevenrsfs
\scriptscriptfont\rsfsfam=\fiversfs
\def\mathscr#1{{\fam\rsfsfam\relax#1}}

\begin{document}

\thispagestyle{empty}

\begin{center}
\begin{small}
\hfill CERN-PH-TH/2005-063 \\
\hfill NEIP/05-05
\end{small}
\begin{center}

\vspace{1.5cm}

{\LARGE \bf D-type supersymmetry breaking\\[2mm] and brane-to-brane gravity mediation
 }

\end{center}

\vspace{1cm}

{\large \bf T.~Gr\'egoire$^{a}$, R.~Rattazzi$^{a}$\footnote{On leave 
from INFN, Pisa, Italy.}, C.~A.~Scrucca$^{b}$\\[2mm] }

\vspace{7mm}

${}^a${\em Physics Department, Theory Division, CERN, CH-1211 Geneva 23, Switzerland}
\vspace{.2cm}

${}^b${\em Institut de Physique, Universit\'e de Neuch\^atel, CH-2000 Neuch\^atel, Switzerland}
\vspace{.2cm}

\end{center}

\vspace{0.4cm}
\centerline{\bf Abstract}
\vspace{-0.1cm}
\begin{quote}
We revisit the issue of gravitational contributions to soft masses in five-dimensional sequestered models. 
We point out that, unlike for the case of $F$-type supersymmetry breaking, for $D$-type breaking these 
effects generically give positive soft masses squared for the sfermions. This drastically improves model building.
We  discuss the phenomenological implications of our result.

\vspace{5pt}
\end{quote}

\newpage

\renewcommand{\theequation}{\thesection.\arabic{equation}}

\section{Introduction}\setcounter{equation}{0}

The generic presence of new sources of Flavor and CP violation in the soft supersymmetry breaking parameters
is one major conceptual drawback lingering on the supersymmetric solution to the hierarchy problem.
This problem was somewhat disregarded in the early models with gravity mediated soft terms with the back 
thought that gravity is a universal force which therefore naturally delivers flavor preserving soft masses. 
On the other hand, whatever theory, perhaps string theory, describes gravity at a more fundamental level, that 
theory should also explain the mass spectrum of quarks and leptons. If quantum gravity is also a theory of flavor 
then sfermion masses are expected to be flavor breaking as well. Mechanisms to control the size of these effects 
have been proposed. 
One idea, more or less satisfactory, is that the same flavor symmetry selection rules that control the hierarchy of 
fermion masses and mixings control the soft terms as well. A perhaps more ambitious direction, using the universal 
character of gravity in the infrared, is to sequester in an extra dimensional space the supersymmetry breaking dynamics 
from the quarks and leptons \cite{Randall:1998uk}. In the simplest case of a single extra dimension, soft masses
are forbidden by locality at the classical level. In this setup, and in the limit of large radius of compactification, the 
soft masses are dominated by the superconformal anomaly contribution \cite{Randall:1998uk,Giudice:1998xp} which 
is nicely flavor preserving, but unfortunately, the sleptons are tachyons. Furthermore, calculable quantum gravity 
corrections to the K\"ahler potential mixing the hidden and the visible sectors give also rise to a universal contribution 
to sfermion masses. For small enough radius this effect competes with anomaly mediation and has the potential to 
cure the slepton mass problem \cite{Randall:1998uk}. Explicit calculations have concluded that this contribution is 
unfortunately always negative \cite{Rattazzi:2003rj, Buchbinder:2003qu}. However it  was also found that yet another 
contribution, due to the radion field $T$, becomes positive in the presence of localized kinetic terms 
with sizable coefficients \cite{Rattazzi:2003rj}. It is in principle possible that the radion contribution dominates the 
sfermion masses, solving the tachyon problem. In practice it turns out to be difficult to construct models where this 
happens, simply because the radion auxiliary Vacuum Expectation Value (VEV) tends to be suppressed \cite{Rattazzi:2003rj,Luty:1999cz}. The more general
situation of a warped extra dimension has also been studied in detail, and it was found that the situation remains 
qualitatively similar \cite{Gregoire:2004nn} (see also \cite{Scrucca:2004cw} for a brief general overview). Finally,
models with several extra dimensions also don't seem to provide any clue to the problem, since 
sequestering is no longer automatic and seems to be lost as soon as important new features are incorporated
\cite{Anisimov,Falkowski:2005zv}.

The purpose of this paper is to point out that the brane-to-brane mediated soft masses, in the basic case 
of a single flat extra dimension, can be made positive simply by choosing a hidden sector in which the order 
parameter of supersymmetry breaking is mostly a gauge auxiliary field $D$ instead of an $F$ auxiliary field \footnote{While we were completing this work, reference  \cite{Falkowski:2005zv} appeared, in which a similar mechanism was used to get positive soft masses at tree level in a six-dimensional context. }.
We will do this in the next section. We will then prove that there is no obstacle in the construction of supersymmetry 
breaking models where $|D| \gg |F|$  (we stress that we are not using Fayet-Iliopoulos terms, which are not acceptable 
at the supergravity level). Finally we will carefully analyze a simple model of radius stabilization and estimate the 
size of all possible subleading flavor breaking effects.

\section{D-type breaking}
\setcounter{equation}{0}

Consider a sequestered model on $S_1/Z_2$ with tree level kinetic function
\be
\Omega= -3M_5^3(T+T^\dagger) + \Omega_{\rm v}(QQ^\dagger) +\Omega_{\rm h}(Xe^{gV} X^\dagger) \,,
\ee
and superpotential
\be
W = W(T) + W_{\rm v}(Q) + W_{\rm h}(X) \,.
\ee
$T$ is the radion superfield and $Q$ and $X$ denote collectively the chiral superfields of respectively 
the visible and hidden sectors \footnote{In what follow, we will denote the scalar component of a chiral multiplet
with the same letter as the corresponding superfield.}. We also indicate by $V$ the gauge superfield localized at the hidden brane,
while we do not explicitly write the visible sector gauge fields because they are irrelevant to the present discussion.  
In general,  $X$ will transform under a reducible representation of  the hidden gauge group $G$.
 Notice that $\Omega$ is related to the K\"ahler potential 
by $K\equiv -3 M_P^2 \ln (-\Omega/3M_P^{-2})$, so that the absence of direct coupling between hidden and visible sectors
is manifest in $\Omega$ but not in $K$. At 1-loop, $\Omega$ is corrected by the additive contribution 
\cite{Rattazzi:2003rj,Buchbinder:2003qu, Gherghetta:2001sa} (see also \cite{Falkowski:2005fm}):
\be
\Omega_{\rm 1-loop}=\frac{\xi(3)}{6\pi^2}\frac{1}{(T+T^\dagger)^2}\Bigl \{\frac{3}{2}
+\frac{\Omega_{\rm v}+\Omega_{\rm h}}{M_5^3(T+T^\dagger)}
+\frac{\Omega_{\rm v}\Omega_{\rm h}}{M_5^6(T+T^\dagger)^2}+\dots
\Bigr \}
\label{1-loop}
\ee
where the dots indicate higher order terms, negligible for our purposes (the full result can be found in 
ref.~\cite{Rattazzi:2003rj}). To discuss soft masses it is enough to consider only the lowest-order term 
in the visible sector kinetic function:
\be
\Omega_{\rm v}=QQ^\dagger \,.
\ee
Neglecting $F_T$, the brane-to-brane contribution to the soft masses coming from the third term in eq.~(\ref{1-loop})
is the dominant effect. If the supersymmetry breaking dynamics in the hidden sector were accurately described by 
weakly coupled chiral superfields, then the resulting universal mass would be
\be
m_0^2 \,=\,-\,\frac{\xi(3)}{6\pi^2M_5^6(T+T^\dagger)^4 }(\partial_X\partial_{X^\dagger} \Omega_{\rm h}) |F_X|^2 
\,<\,0 \,,
\label{tachyon}
\ee
where positivity of $\partial_X\partial_{X^\dagger} \Omega_{\rm h}$ has been used. 
The situation changes however when the gauge dynamics in the hidden sector cannot be neglected. Basically
this means that the vector fields are light and the corresponding auxiliary fields have non-negligible VEVs. 
Without loss of generality we can limit ourselves to the simple case of a quadratic, renormalizable, kinetic term
\be
\Omega_{\rm h} = X^\dagger e^{g V} X \,.
\ee
The soft mass $m_0^2$ is determined by the VEV of the highest component of the superfield $X^\dagger e^{gV} X$.
Working in Wess-Zumino gauge this reduces to
\bea
X^\dagger e^{gV} X|_{\theta^2\bar \theta^2} \!\!\!&=&\!\!\! |F_X|^2 + g D_A X^\dagger T_A X
= |\partial_X W_{\rm h}|^2 -g^2 (X^\dagger T_A X)^2 \nn \\
 \!\!\!&\equiv&\!\!\! |F_X|^2-D^2 \,,
\eea
where in the second equality the equations of motion for $F_X$ and $D$ have been used. A supersymmetry
breaking model with $|D| > |F_X|$ would induce a universal positive contribution to the scalar masses, potentially solving the tachyon problem of anomaly mediation. 
While in the scalar potential $V=|F_X|^2+D^2/2$ the contributions of the two auxiliary fields add up, in the brane-to-brane mass they enter with opposite signs. This "asymmetric" behavior of $F$ and $D$ arises because $X^\dagger e^{gV} X|_{\theta^2\bar \theta^2}$ contains the "kinetic" term for the first and the "source" term for the second. On the other hand the kinetic term for $D$ comes from the hidden gauge superfield 
strength $W_\alpha W^\alpha |_{\theta^2}$ while the "source" term for $F$ is due
to the hidden superpotential. However, based solely on power counting, these latter interactions are found to be unimportant 
for gravity mediated masses. For example, along with the terms in eq.~(\ref{1-loop}),
bulk gravity loops could introduce a visible-hidden mixing of the form $
[Q Q^\dagger W_\alpha W^\alpha/M_5^6(T+T^\dagger)^3]_{\theta^2\bar \theta^2}$. 
Notice that this is technically a higher derivative interaction  than the K\"ahler potential. As a result, the induced soft mass is suppressed by at least an extra factor $m_{3/2} T$ with respect to the leading contribution in eq.~(\ref{tachyon}). Concerning the "source" term for F, one could imagine that similar terms, connecting the visible kinetic term and the hidden sector superpotential, arise. Again, since the superpotential has higher dimension than the K\"ahler potential, these terms will give soft masses that are suppressed compare to the leading contribution. For example, an operator of the form $[Q^\dagger Q W_{\rm h}(X) /M_5^6 (T+T)^3|]_{\theta^2 \bar \theta^2}$ would also give a contribution to soft masses suppressed by $m_{3/2} T$ (notice that $W_{\rm h}|_{\theta^2} = |F_X|^2$). At the leading
order it is therefore consistent to consider only the effects of the matter kinetic term
$X^\dagger e^{gV} X$.

Even though in principle there does not seem
to be anything wrong in having a model with $|D|>|F_X| $, we will nonetheless show in the next section that this is 
indeed the case by presenting a model in which $|D| \gg |F_X|$.  Before going to that, it would be useful 
to give the {\it on-shell} viewpoint on the above result. The coupling between hidden and visible sectors can also be 
treated as a correction to the kinetic function of the hidden sector, by defining
\be
\tilde \Omega_{\rm h} \equiv  \left (1+ a\,Q^\dagger Q\right ) X^\dagger e^{gV} X\,,
\ee
with $a=(\zeta(3)/6 \pi^2) M_5^{-6} (T + T^\dagger)^{-4}$. Neglecting the supergravity corrections, which 
contribute at a higher order in $XX^\dagger/M_P^2$, and the effects associated to $F_T$,
the hidden sector potential is  found to be
\bea
V_{\rm h} \!\!\!&=&\!\!\!\frac{|\partial_X W_{\rm h}|^2}{\partial_X\partial_{X^\dagger} \tilde \Omega_{\rm h}}
+\frac{g^2}{8}\left (\partial_X \tilde \Omega_{\rm h} T_A X +{\rm h.c.}\right )^2 \nn \\
\!\!\!&=&\!\!\!\frac{|\partial_X W_{\rm h}|^2}{1+ a\, Q^\dagger Q} 
+ \frac{g^2}{2}(1+ a\, Q^\dagger Q )^2 (X^\dagger T_A X)^2 \;.
\eea
This equation shows that the fact that the two contributions to the mass of $Q$ from respectively $F_X$ 
and $D$ terms have opposite signs is simply due to the different powers of $\tilde \Omega_{\rm h}$ appearing in 
these two terms. The new expression for the scalar soft mass is then:
\be
m_0^2 \,=\, \frac{\xi(3)}{6\pi^2M_5^6(T+T^\dagger)^4 } 
\Big(- |F_X|^2 +  D^2 \Big) \,.
\label{notachyon}
\ee

\section{$D$ versus $F$ auxiliary fields}
\setcounter{equation}{0}

Consider a globally supersymmetric theory with matter field $X$ belonging to a generally reducible representation of the gauge group $G$ and with 
scalar potential
\be
V= |\partial _X W|^2+ \frac{g^2}{2} (X^\dagger T_A X)^2 \,.
\ee
There is a well know theorem stating that if a solution $F_X=\partial _X W=0$ exists, then there must also exist a 
value of $X$ satisfying both conditions $F_X=D_A=0. $\footnote{See ref.~\cite{wessbagger}. 
This theorem can be considered a corollary of another theorem, which states that the space of $D$-flat directions 
is isomorphic to the space of holomorphic gauge invariants  \cite{Buccella:1982nx, Procesi:1985hr, Gatto:1985jz, Luty:1995sd}.}
At first sight, this result suggests that there may be an obstruction in constructing models with $|D_A|$ bigger that 
$|F_X|$. More precisely, the case in which $|F_X|$ is much smaller than $|D_A|$ approximates well the case $F_X=0$, suggesting that starting from a point satisfying this condition, $D_A$ can be reduced without significantly affecting $F_X$.
There is however a caveat in this simple reasoning, and, as we will show below, it is  possible to find models with 
$|D_A|/|F_X|$ arbitrarily large. Of course, in order to make the gravity mediated masses positive we do not need that 
much, but the fact that it is possible suggests to us that the case we care about, $|D_A|^2> |F_X|^2$, may be pretty generic. 
In analogy with the standard proof of the theorem stated above, let us start from a point $X=X_0$ in field space, and 
consider a particular variation of the fields corresponding to a complex gauge rotation:
\be
X_0 \to e^{\alpha_A T_A} X_0 \;,\quad \alpha_A  \; \mbox{real}\, .
\ee
At first order in $\alpha_A$ the potential will change by
\bea
\nonumber
\delta V \!\!\!&=&\!\!\!  \alpha_A \Big\{\left [ (\partial_X  W)^\dagger T_A (\partial_X W) + {\rm h.c.}\right ]_{X=X_0}
+ g^2 (X^\dagger_0 \{T_A,T_B\} X_0) (X_0^\dagger T_B X_0)  \Big\} \nn \\
\!\!\!&=&\!\!\! \alpha_A \Big\{2 F_X^{\dagger} T_{A} F_X + M_{AB}^2 D_B \Big\} \,,
\eea
where $M_{AB}$ is the mass matrix of the heavy gauge bosons, and we have used the fact that $\partial _X W$ 
transforms under complex gauge transformations as the conjugate of $X$. Consider first the case where $F_X=0$ at 
$X_0$. We indicate by $H$ the subgroup of $G$ that leaves this point invariant. The $D$ field transforms in the 
adjoint representation of the gauge group, and its VEV at $X_0$ is along the subspace $G/H$. Moreover, the vector mass 
matrix is a non-singular matrix on the subspace $G/H$. It is then obvious that for $D \not = 0$ the variation of $V$ 
is in general non-zero. Therefore points with $D\not = 0$ cannot be stationary. 
Consider now the case where $F_X \neq 0$. At the minimum,  $\delta V=0$ and thus
\be
2 F_X^\dagger T_A F_X = - M_{AB}^2 D_B \,.
\label{stationary}
 \ee
Now, if the Lie algebra charges are $O(1)$ parameters, the above equality roughly implies
\be
|F_X|^2 \gsim 2 |F_X^\dagger T_{A} F_X| = M_{AB}^2 D_B \gsim  D^2 \,,
\ee
showing that $D^2$ can conceivably be of order $|F_X|^2$ but not much bigger. The latter possibility can however be achieved in the presence of large parameters in the charge spectrum. Consider for example a simple $U(1)$ model 
with matter fields $X_{q_i}$ with charges $q_i$. Imagine now that the dominant $F$-auxiliary at the stationary point 
belongs to a superfield $X_{q_k}$ with charge $q_k\equiv N \gg 1$, whereas all the scalar (non-auxiliary) fields with 
non-zero VEV belong to superfields with {\it small} charge, say $O(1)$. 
In such a situation the left hand side of eq.~(\ref{stationary}) is of order $N |F_X|^2$, while the right hand side can 
conceivably be of order $D^2$. In this way we can parametrically obtain $D^2 \gg |F_X|^2$.
Once again, we emphasize that for our application we do not need this parametric separation, but only that 
$- |F_X|^2 + D^2$ becomes positive and of the same order of magnitude as $|F_X|^2$ in eq.~(\ref{notachyon}).

We can make the above more explicit by considering the following O'Raifertaigh model with a gauged $U(1)$. We introduce 4 chiral superfields $\phi_0,\phi_1,\phi_{-1},\phi_{-1/N}$ with charges $0,1,-1,-1/N$ respectively and superpotential
\be
W = \lambda_1\, \phi_0(\phi_1\phi_{-1/N}^N-1) + \lambda_2\, \phi_1 \phi_{-1}\, .
\ee
For the moment we shall not worry about  the above charge assignment being anomalous. 
By defining the invariants
\be
I_1\equiv \phi_1\phi_{-1/N}^N \;,\qquad I_2\equiv \phi_{-1} \phi_{-1/N}^{-N} \,,
\ee 
the scalar potential is
\bea
V \!\!\!&=&\!\!\! |\lambda_1|^2|I_1-1|^2 + |\lambda_1 \phi_0 + \lambda_2 I_2|^2 |\phi_{-1/N}|^{2N}
+ |N \lambda_1\phi_0 I_1|^2 |\phi_{-1/N}|^{-2} \\
\!\!\!&\;&\!\!\! +\,|\lambda_2 I_1|^2 |\phi_{-1/N}|^{-2N} + \frac{g^2}{2} \Big(|I_1|^2 |\phi_{-1/N}|^{-2N} - |I_2|^2 |\phi_{-1/N}|^{2N}
- \frac{1}{N} |\phi_{-1/N}|^2\Big)^2\,. \nn
 \eea
To further simplify notation, we define $|\phi_{-1/N}|^2\equiv z$, $|I_2|^2= w$, $I_1\equiv y e^{i\theta}$. 
By extremizing in $\phi_0$ and $\theta$, we are left with finding the minimum of
\bea
V \!\!\!&=&\!\!\! |\lambda_1|^2(y-1)^2 + |\lambda_2|^2 \frac{N^2y^2 z^N w}{N^2y^2+z^{N+1}} 
+ |\lambda_2|^2\frac{y^2}{z^{N}} \nn \\
\!\!\!&\;&\!\!\! +\, \frac{g^2}{2}\left (\frac{y^2}{z^N}-w z^{N} -\frac{1}{N}z\right )^2 \,.
\eea

In the limit of zero gauge coupling, there is an asymptotic supersymmetric minimum at $w=0,y=1$ and 
$z \rightarrow \infty$. When the gauge coupling is turned on, the vacuum moves in from infinity to settle 
at some finite value of $z$, and supersymmetry is broken. To study the vacuum dynamics, it is convenient 
to define the rescaled couplings $\hat \lambda_i\equiv \lambda_i/g$. We can simplify the minimization of 
the potential  by taking  $\hat \lambda_1\gg \hat \lambda_2 \gg 1$. Then  we have that $y\sim 1+O(1/\hat \lambda_1^2)$ 
and $F_{\phi_0}= \hat \lambda_1(y-1) \sim 1/\hat\lambda_1$ is negligible. It is also easy to guess that the 
minimum should lie at $w=0$. We  can then consistently approximate the potential  by
\be
V/g^2\simeq \hat\lambda_2^2\frac{1}{z^{N}}+\frac{1}{2} \frac{z^2}{N^2} \,.
\label{Ndep}
\ee
This is stationary for $z^{N+2}\sim N^3\hat\lambda_2^2\gg 1$,
and at the stationary point the vacuum energy scales like
\be
V_{\rm min}\sim g^2 (N^3\hat\lambda_2^2)^{\frac{2}{N+2}}\left (\frac{1}{N^3}+\frac{1}{2N^2}\right ) \,.
\ee
The $F$ and $D$ contributions correspond respectively to the first and second term in the parentheses. Notice that  $D^2/|F|^2\sim N$ as promised in the qualitative discussion above: the $ |F|^2$ term in eq.~(\ref{Ndep}) is associated to the
auxiliary of a field of charge $-1$, while the $D^2$ term is associated to the VEV of a field of charge $-1/N$.

To get an anomaly free model, it is sufficient to add two additional superfields, $\phi_0'$ and $\phi_{1/N}$, with charge $0$ 
and $1/N$ respectively. This makes the model manifestly vector-like. To qualitatively preserve the characteristics of the 
minimal model, one can then add the following term to the superpotential:
\be
\Delta W = \lambda_3\, \phi_0' \phi_{1/N} \phi_{-1/N} \,.
\ee
It is clear that for large enough $\hat \lambda_3$, this will force $\phi_0' = \phi_{1/N}=0$, thus leaving  the minimization of the 
potential  unaffected.

Note that the above model serves only as an illustration for the possibility of obtaining a large $D$-type auxiliary field 
in a perturbative  setup. Models of dynamical supersymmetry breaking can also give rise to a non-vanishing $D$-type
auxiliary field, and a priori there is no reason to expect the VEVs of vector multiplet auxiliary fields to be much smaller 
than the VEVs of chiral multiplet auxiliary fields. For instance, we examined  the $4-1$ model of 
ref.~\cite{Dine:1995ag} (see also \cite{Carpenter:2005tz} for the use of such model to get significant $D$ auxiliary 
field). This model can be studied numerically, and we found regions of parameter space where $|D| \gsim |F|$ (see also \cite{Carpenter:2005tz}).  

In conclusion we have proven that, when  the supersymmetry breaking dynamics is described by a set  of {\it weakly} coupled superfields (composite or elementary),
the VEV $\langle X^\dagger e^{gV} X|_{\theta^2\bar \theta^2}\rangle$ is {\it not} positive definite. At weak coupling, this quantity becomes negative when
$D$-auxiliary fields are big enough. It is reasonable to conclude that it may well be negative also in models that
break supersymmetry in the strongly coupled regime (for which there is no weakly coupled description, elementary or composite). An example in
such a class is given by an $SU(5)$ gauge theory with matter in a $10 +\bar 5\equiv X$ representation \cite{Affleck:1983vc,Meurice:1984ai}.
Now $ X^\dagger e^{g V} X|_{\theta^2\bar \theta^2}$ is a composite operator whose VEV does not have an obvious sign. A truly non-perturbative
treatment, the lattice, would be needed to infer the sign. But it seems reasonable to conclude that models in this class have an equally good chance to give positive or negative sfermion masses.
The only class of models for which $\langle X^\dagger e^{g V} X|_{\theta^2\bar \theta^2}\rangle $ is always positive are weakly coupled O'Raifertaigh models,
which are just the simplest to deal with. In all other models we expect the sign not to be definite.

\section{Model building and phenomenology}
\setcounter{equation}{0}

We  now discuss the implications of our simple remark.
Almost all the results of this section have appeared before (mostly 
in ref.~\cite{Luty:1999cz}, but also in refs.~\cite{Rattazzi:2003rj,Gregoire:2004nn}) and the purpose of this section 
is to make a synthesis of the important issues in our setup. Let us start by studying the conditions under which 
anomaly mediation (AM) and brane-to-brane loops (B2B) contribute in a comparable way to the sfermion masses. 
We first have to relate the compensator VEV $F_\phi$ to the scale $M_S$ of supersymmetry breaking in the hidden 
sector. This is done by demanding that the effective four-dimensional cosmological constant vanishes. The hidden 
sector gives a positive contribution $V_{\rm hid}=|F|^2+|D|^2/2 \sim M_S^4$, while the bulk gravity sector, after radius stabilization\footnote{Later in the section, we will illustrate this by considering the model of ref.~\cite{Luty:1999cz}}, 
gives a negative contribution $V_{\rm bulk} $ roughly equal to $-M_5^3(T+T^\dagger)|F_\phi |^2$. Therefore by
demanding $V_{\rm hid}+V_{\rm bulk}=0$ we obtain
\be
| F_\phi |^2 \sim \frac {|F|^2 +D^2/2}{M_5^3 T}=\frac{|F|^2+D^2/2}{M_P^2}\,.
\ee
In the end, this just corresponds to the usual relation $M_S^4\sim M_P^2 m_{3/2}^2$. By eq.~(\ref{1-loop}),
equality of orders of magnitude of AM and B2B then corresponds to
\be
\left (\frac{\alpha}{4 \pi}\right )^2\sim \frac{1}{16 \pi^2 (M_5 T)^3}\equiv \alpha_5 \,.
\label{equality}
\ee
Notice that the quantity $\alpha_5$ corresponds to the loop expansion parameter controlling gravitational
quantum corrections at the compactification scale. Numerically, using $M_P^2=M_5^3T=10^{18}{\rm GeV}$ 
the above relation implies a compactification radius $1/T\sim 10^{17} {\rm GeV}$, slightly above the GUT scale.

With the size of $\alpha_5$ fixed, we can estimate the size of flavor violation coming from higher order gravitational interactions. The allowed operators are most conveniently classified by using the superfield formulation of 
linearized supergravity  \cite{Buchbinder:1998qv,Gates:1983nr,Linch:2002wg} (see also 
\cite{Buchbinder:2003qu,Gregoire:2004nn}). On the branes, for example, it is possible to have a four-derivative
operator of the form
\be
{\cal L}^{\partial^4} \sim \int d^4 \theta\, \phi^\dagger \phi\,  \frac{Z_{ij}}{\Lambda_5^2} Q_i^\dagger Q_j (K_{mn} V^n)^2 \,.
\label{4derivatives}
\ee
The tensor $K_{mn}$ is the two-derivative kinetic operator for supergravity, $V_n$ is the supergravity superfield 
containing the $4D$ part of the graviton and the gravitino, and $\phi$ is the conformal compensator chiral multiplet. 
The scale $\Lambda_5^3 \sim 16\pi^2 M_5^3$ is the NDA cut-off of five-dimensional gravity\footnote{We 
are following the remark made in \cite{Agashe:2005vg} that this is a more accurate estimate of the NDA scale than the 
usually used one $24 \pi^3 M_5^3$. In the end, the extra ``floating'' $\pi$ does not make a big difference in our estimates.}. We have checked that this operator can be fully covariantized. The quantity $Z_{ij}$ is a matrix in flavor space, which 
differs in general from the two-derivative kinetic matrix (or equivalently, it differs from the identity $\delta_{ij}$ in the 
basis where the two-derivative term is canonical) and adds a new source of flavor violation, different from the usual Yukawa matrices. From a gravitational viewpoint, the presence of these extra terms
corresponds to tiny violations of the equivalence principle at the quantum gravity scale: two different quarks, say the 
top and the charm, will follow different trajectories in an external field. From the viewpoint of flavor physics, on the other 
hand, these terms are like any other violation of the GIM mechanism. The latter, when valid, is an accidental property 
of the truncation of the Lagrangian to the lowest-order operators in the derivative expansion. In the absence of extra 
flavor selection rules, NDA suggests that $Z_{ij} \sim 1$. This would corresponds to maximal flavor violation happening 
in processes at the quantum gravity scale $\Lambda_5$. Inserting the operators of eq.~(\ref{4derivatives}) into the graviton loop mediating the B2B effects can produce flavor-violating soft masses that can be estimated by noticing that the loop is dominated at virtuality of order $1/T$. We find the relative size of flavor breaking and flavor preserving B2B masses to be:
\be
\frac{\delta m_{ij}^2}{m_0^2} \sim \frac{1}{\Lambda_5^2 T^2} = \alpha_5^{\frac{2}{3}}\sim 
\left(\frac{\alpha}{4\pi} \right )^{\frac{4}{3}}\sim 10^{-3}\,.
\label{bound}
\ee

In principle there could also be three-derivative flavor-violating operators similar to eq.~(\ref{4derivatives}), containing 
two $V_n$ fields and superspace derivatives, with a structure of the form $Q_i^\dagger Q_j V \partial \partial D^2 V$. 
Such an operator, however, contains only $D$ and no $\bar D$, and is thus charged under $R$-symmetry. Its contribution to soft masses would 
thus either be suppressed by an extra power of $m_{3/2}/M_P$, or,  if it was combined with a similar operator of 
opposite $R$-charge, give an effect comparable to that of the operator in eq.~(\ref{4derivatives}).

We can compare the above result  with the present experimental bounds on flavor violation in the MSSM. The strongest bounds
are associated to $\epsilon_K$ and to ${\rm Br}(\mu\to e\gamma)$.  Assuming $O(1)$ phases, the bound on $\epsilon_K$ implies
\be
\left (\frac{\delta m^2}{m^2}\right )^2 \lsim 10^{-7}\left (\frac{m}{300 {\rm GeV}}\right )^2 \,,
\ee
which compared to our result eq. (\ref{bound}), requires either some extra mild suppression of flavor breaking in  $Z_{ij}$ or squarks heavier than a TeV.  The first possibility is, for instance, comfortably realized in all flavor models
based on the Froggat-Nielsen mechanism, where we expect a further suppression of $\delta m^2_{12}$ by at least one
power of the Cabibbo angle $\theta_C \simeq 0.2$. In practice, however, the second possibility is ``unfortunately'' almost
likely forced on us by the bounds on the SUSY and Higgs particle masses after LEP. A detailed analysis of electroweak
breaking would however be needed to make a more precise statement. In one or the other way, it seems rather easy to
satisfy the bound on $\epsilon_K$, although it is possible that the supersymmetric contribution amounts to a sizable 
fraction of the observed value of $\epsilon_K$.  The other strong bound comes from ${\rm Br}(\mu\to e\gamma)$.  There are various diagrams contributing to this
process in the MSSM. Focussing on the chargino induced contribution, which is 
dominant (or at least not subleading) in a significant variety of situations, and assuming 
charginos and sleptons of comparable mass, we obtain the rough 
 estimate  (see for instance \cite{Hisano:1995cp,Rattazzi:1995ts}):
\be
{\rm Br}(\mu\to e\gamma)\sim 5 \times 10^{-11} \left (\frac{\delta m^2/m^2}{0.001}\right )^2 
\left (\frac{150 {\rm GeV}}{m}\right )^4 \, .
\label{Br}
\ee
This is right at the boundary of the experimentally allowed region and significatively above the planned precision $\sim 10^{-13}$ 
of the ongoing experiment at PSI \cite{Baldini:2004dj}, unless further sizable suppressions from flavor selection rules intervene. 

If the flavor physics scale is the cut off $\Lambda_5$ then the flavor violating effects we discussed above  are the only ones we expect.
This is a consistent but restrictive  situation, and it is worth considering the perhaps more realistic situations where
there is some flavor dynamics at lower scales.

One such situation occurs if the Standard Model gauge group is unified at a scale $\sim 10^{16}$ GeV. Since
quarks and leptons become also unified, the flavor violation encoded in the large top Yukawa coupling can generate flavor violation 
in the lepton sector through renormalization group evolution between the scale at which the soft masses are generated 
and the GUT scale. This effect was considered in refs.~\cite{Hall:1985dx,Barbieri:1995tw}, where  universality of soft
masses was assumed to hold at the Planck scale and the effect on ${\rm Br}(\mu\to e\gamma)$ was found to be close to 
the bound.  In our case, the 
soft masses are universal at $1/T$, therefore there is less running to the GUT scale, and we expect the effect to be smaller. Moreover, universal $A$ terms
at the cut-off scale  were also assumed in \cite{Hall:1985dx,Barbieri:1995tw}, so that running above $M_{\rm GUT}$ induced flavor violation in the leptonic A-terms.
On the contrary, in our setup, $A$ terms exactly follow their AMSB trajectory, and therefore high energy flavor violation
decouples from their low energy value \cite{Randall:1998uk}. Therefore, in our case, lepton flavor violation only resides in the dimension 2 slepton masses. It can  be estimated 
by running  the third-family soft mass from the compactification scale $1/T$ to the GUT scale $M_{\rm GUT}$. This effect 
depends on the value of the top Yukawa coupling at the GUT scale and is typically enhanced by a large group-theory
factor. In the case of a minimal $SO(10)$ theory, for example, one finds the following soft mass splitting between the third 
and the first two generations:
\be
\frac {\delta m_{3}^2}{m^2} \simeq \frac {15}{8 \pi^2} \lambda_t^2 \ln (M_{\rm GUT} T) 
\simeq 0.3 \left(\frac {\lambda_t}{0.8} \right)^2 \,.
\label{dm3}
\ee
$SU(5)$ does not lead to bigger effects.
After rotating the matter superfields to diagonalize the Yukawa couplings, this induces a soft mass 
mixing between the first two generations. In general, even for a fully realistic theory of flavor at the GUT scale we expect the rotation
matrix $V_\ell$ in the lepton sector to have entries comparable to those of the CKM matrix $V$ \footnote{For instance in the naive minimal $SU(5)$ model with
renormalizable Yukawa interactions, this $V_\ell$ is exactly the CKM matrix, but at the same time one also gets the wrong predictions $\lambda_e=\lambda_d$, $\lambda_\mu=
\lambda_s$ at the GUT scale.}. We then obtain for the 1-2 entry of the slepton mass matrix the rough estimate
\be
\frac {\delta m^2}{m^2} \sim V_{13} V_{23} \frac {\delta m_{3}^2}{m^2} 
\simeq  10^{-4} \left(\frac {\lambda_t}{0.8} \right)^2 \,.
\label{dm}
\ee
Neglecting the running down to the weak scale, which  does not qualitatively change the result, and using eq.~(\ref{Br}) we find
\be
{\rm Br}(\mu\to e\gamma)\sim 5 \times 10^{-13}  \left(\frac{\lambda_t}{0.8} \right)^4   
\left (\frac{150 {\rm GeV}}{m}\right )^4 \, ,
\label{so10}
\ee
which is below the present bound, but again within reach of future experiments \footnote{In the case of $SO(10)$, as noticed in ref. Ò\cite{Barbieri:1995tw}, the presence of mixing in both right-handed and left handed sleptons makes another diagram, involving $A-$terms and enhanced  by a factor
$m_\tau/m_\mu$, important. We have checked, that because of other suppressions
(this diagram involves the smaller hypercharge coupling  and needs a double insertion of the mass splitting  in eq.~(\ref{dm3})) our estimate eq.~(\ref{so10})
is not qualitatively modified. A more detailed analysis, accounting for the
specific features of our spectrum, is clearly needed in order to make more precise statements.}.

Another likely source of lepton flavor violation, in principle independent of unification, is associated to the neutrino Yukawa couplings.
If neutrinos are Dirac particles, then these Yukawa couplings are so small that their effect is totally negligible. However, if, as it seems more plausible,
neutrinos turn out to be light because of the see-saw mechanism, then they could have sizable and even $O(1)$ Yukawas. This biggest value
would correspond to a right handed neutrino mass
around $ \sim 10^{14}$ GeV. The induced effect is analogous to the one just discussed in Grand Unified models: Yukawa flavor violation feeds into
the slepton masses via RG evolution from the scale $1/T$ or $M_{\rm GUT}$ down to the relevant right-handed neutrino mass $M_N$, at which the theory flows to the 
MSSM \cite{Borzumati:1986qx}. The relevance of this effect however strongly relies on the size of the neutrino Yukawa, which unlike $\lambda_t$ in the previous discussion, is not fixed directly by experiments, and may well be, say, $O(10^{-2})$ in which case the effects would not be dramatic.
The only case in which we necessarily expect some sizable neutrino Yukawa is for
GUTs, like $SO(10)$ or bigger groups,  where neutrinos and quarks are unified. In this case however the mixings
due to neutrino Yukawa matrices are favored to be CKM-like, with the large lepton flavor mixing angles measured at low energy coming from the interplay between the Yukawa and the right-handed neutrino Majorana mass matrix.
Then in these  models we expect extra contributions that are similar, but not bigger than the estimate in eq.~(\ref{dm}).
Therefore we conclude that, while it is possible that neutrino induced lepton flavor violation becomes quite large, there is no necessary reason for it to be the case \cite{Masiero:2004js}.

To summarize, we have identified at least 3 sources of flavor violation in our scenario that could contribute in an important way to the rate for $\mu\to e\gamma$. While all these effects are more or less model dependent, they seem both small enough to be safely compatible with the present bound and big enough to raise hopes for a signal in the ongoing experimental search for $\mu \to e \gamma$.

To conclude this section, we focus on a simple explicit model of radius stabilization
\cite{Luty:1999cz} to illustrate that there needn't exist flavor breaking corrections bigger than the ones we have 
discussed and coming from the radius stabilization mechanism. The model involves two Super Yang-Mills sectors, one 
on the hidden brane and one in the bulk. Gaugino condensation in the low energy effective theory leads to a 
superpotential of the form
\be
W =\frac{1}{16\pi^2}\bigl ( \Lambda_1^3 +\Lambda_2^3 e^{-a\Lambda_2 T}\bigr ) \,,
\ee
where $\Lambda_{1,2}$ are the strong coupling scales of respectively the boundary and bulk gauge interactions, 
while $a$ is an $O(1)$ coefficient. As emphasized in ref.~\cite{Luty:1999cz}, at the leading order in an expansion 
in $1/T\propto 1/M_P^2$, it is consistent to study the radius potential by neglecting the supersymmetry breaking 
sector localized at the hidden brane. The first remark is that, apart from the Goldstino, all the fields in that sector 
can conceivably have a mass of order $M_S \gg m_{3/2}$, and they can be integrated out before studying radius stabilization. Secondly, as it can be directly checked, the supersymmetry breaking corrections to the radion potential induced by integrating out the hidden sector are controlled by the gravitino mass and are parametrically smaller
than the potential induced by $W$. At the minimum of the radion potential, we obtain
\bea
\label{radstab1}
\Lambda_2T \!\!\!&\simeq&\!\!\! 3\ln (\Lambda_2/\Lambda_1) +\ln(\Lambda_2 T) \,,
\label{radion} \\
\label{radstab2}
F_\phi \!\!\!&\simeq&\!\!\! \frac{\Lambda_1^3}{16\pi^2 M_5^3 T}\simeq m_{3/2} \,,\\ 
\frac{F_T}{T} \!\!\!&\simeq&\!\!\! \frac{m_{3/2}}{\Lambda_2 T}\, ,\\
m_T \!\!\!&\simeq&\!\!\! m_{3/2} (\Lambda_2 T) \,.
\eea
Notice that in the perturbative limit, where $\Lambda_2 T \gg 1$, we obtain $m_T \gg m_{3/2}$ and $F_T/T \ll m_{3/2}$, 
which is consistent with neglecting gravity mediated effects associated to the radion, as just discussed. A 
reasonable assumption, that allows us to  proceed further, is that the quantum scales of the SYM theory and 
gravity in the bulk coincide: $\Lambda_5 \sim \Lambda_2$. In this case,  by using eqs.~(\ref{radstab1}) and  (\ref{radstab2})  we can simply express the gravitational 
expansion parameter as
\be
\alpha_5 \equiv \frac{1}{(\Lambda_5T)^3}\sim \left(\ln \frac {M_P}{m_{3/2}}\right)^{-3} \,.
\ee
As remarked in ref. \cite{Luty:1999cz}, by inputting $m_{3/2}\sim 10^5$ GeV, and for 
$M_P\sim 10^{18}$ GeV, we automatically obtain $\alpha_5 \sim (1/30)^3\sim 10^{-4}$, in 
such a way that AM and B2B are effortlessly of the same order. 

Let us now consider  the potentially dangerous effects mediated by the bulk gauge fields 
in the above model. Since the observable and hidden sector fields are not charged under the bulk gauge group, the relevant  couplings between fields localized on the branes and bulk gauge fields are non-renormalizable interactions of the form
\begin{eqnarray}
\Omega_{\rm v} \!\!\!&\supset&\!\!\! \bigg((A_{\rm v})_{ij}\, \frac {W^\alpha W_\alpha}{\Lambda_2^3}
+ (B_{\rm v})_{ij}\, \frac {D^\alpha W^\beta D_\alpha W_\beta}{\Lambda_2^4} + \cdots + {\rm h.c.}\bigg)\, 
Q^{i\dagger} Q^j \,, \label{AB} \\
\Omega_{\rm h} \!\!\!&\supset&\!\!\! \bigg(A_{\rm h}\, \frac {W^\alpha W_\alpha}{\Lambda_2^3}
+ B_{\rm h}\, \frac {D^\alpha W^\beta D_\alpha W_\beta}{\Lambda_2^4} + \cdots + {\rm h.c.}\bigg)\, 
X^\dagger e^{g V} X \,. \label{CD}
\end{eqnarray}
The quantities $(A_{\rm v})_{ij}$ and $(B_{\rm v})_{ij}$ are a priori generic flavour non-universal matrices, whereas 
$A_{\rm h}$ and $B_{\rm h}$ are numbers, and all of them are expected by NDA to be $O(1)$. Notice however 
that the leading first term in each parenthesis violates $R$-symmetry, and its coefficient could therefore be 
further suppressed compared to the one of the second term in each parenthesis, which preserves $R$-symmetry, 
depending on the extent to which $R$ symmetry is broken. 
The interactions in $\Omega_{\rm v}$ and $\Omega_{\rm h}$, through the exchange of the bulk gauge fields, give rise to potentially
dangerous  flavour violating contributions to sfermion masses.
(Notice that since $\Lambda_2 T \sim 30$, any non-perturbative effect associated to gaugino condensation is 
highly suppressed, and can therefore be neglected). The leading contribution
comes from a one-loop diagram involving a vertex proportional to $A_{\rm v}/\Lambda_2^3$ and one 
proportional to $A_{\rm h}/\Lambda_2^3$. As usual, the loop integral is saturated at momenta of the order of 
the compactification scale $1/T$, and by dimensional analysis it yields a factor of order $g_5^4/(4 \pi^2 T^6)$. 
Since $\Lambda_2 = 8 \pi^2/g_5^2$, this can be rewritten as $16 \pi^2/(\Lambda_2^2 T^6)$. This results finally 
in the following effective operator:
\be
\Delta \Omega_{\rm 1-loop}^{\rm gau} \sim (A_{\rm v})_{ij} A_{\rm h} \frac {16 \pi^2}{\Lambda_2^8 T^6} \,
Q^{i\dagger} Q^j\, X^\dagger e^{g V} X + {\rm h.c.}\,.
\label{op1}
\ee
This has to be compared with the gravitational effect encoded in the third term of eq.~(\ref{1-loop}), which 
can be rewritten as
\be
\Delta \Omega_{\rm 1-loop}^{\rm gra} \sim \delta_{ij} \frac {16 \pi^2}{\Lambda_2^6 T^4} \,
Q^{i\dagger} Q^j\, X^\dagger e^{g V} X \,.
\label{opgrav}
\ee
We see that the effect (\ref{op1}) is down by at least a factor  $1/(\Lambda_2 T)^2$ compared to (\ref{opgrav}),
even in the limit where $R$ symmetry is maximally broken. Since $\Lambda_2 \sim \Lambda_5$, this effect has therefore 
at worse the same size as the one due to subleading gravitational effects. It is straightforward to verify that the effects 
induced by the subleading $R$-preserving couplings in eqs.~(\ref{AB}) and (\ref{CD}) are further suppressed by extra 
powers of $1/(\Lambda_2 T)^2$ and are therefore less important than the subleading gravitational effect. The typical size of flavour violating effects remains 
therefore the one of eq.~(\ref{bound}).

\section{Conclusions}
\setcounter{equation}{0}

We have shown that when $D$-type auxiliaries in the hidden sector have a big enough VEV, then the universal soft scalar masses squared induced by gravitational 
loops  in five-dimensional sequestered models is  positive. We have argued, by
showing two explicit examples, that this situation is not hard to achieve and that it may be pretty generic. This opens up the possibility to build viable models where satisfactory gaugino and sfermion masses are 
naturally generated by a combination of anomaly mediation and brane-to-brane gravity loops. We have also estimated the flavor non-universal subleading corrections that are 
generically expected to occur and found that they do not contradict to present experimental bounds. Moreover we found that the rate for $\mu \to e \gamma$ could plausibly be within the reach of the planned sensitivity of the ongoing experiment.
An issue that remains to be studied  concerns 
the generation of $\mu$ and $B \mu$,
which is known to be a generic weakness of the anomaly mediated scenario,
and the  breaking of the electroweak symmetry. One interesting and perhaps natural direction to explore would be to extend the model by adding a light singlet (NMSSM). We hope to do that in  future work.

\section*{Acknowledgments}

We thank N.~Arkani-Hamed, R.~Barbieri, J.-P.~Derendinger, G.~Giudice, M.~Luty, A.~Strumia and R.~Sundrum for useful discussions. 
This work has been partly supported by the Swiss National Science Foundation and by the European Commission 
under contracts MRTN-CT-2004-005104 and MRTN-CT-2004-503369.
C. S. thanks the Theory Division of CERN for hospitality.


\begin{thebibliography}{10}

\bibitem{Randall:1998uk}
L.~Randall and R.~Sundrum,
Nucl. Phys. {\bf B557}, 79 (1999), hep-th/9810155.
%%CITATION = HEP-TH 9810155;%%

\bibitem{Giudice:1998xp}
G.~F. Giudice, M.~A. Luty, H.~Murayama, and R.~Rattazzi,
JHEP {\bf 12}, 027 (1998), hep-ph/9810442.
%%CITATION = HEP-PH 9810442;%%

\bibitem{Rattazzi:2003rj}
R.~Rattazzi, C.~A. Scrucca, and A.~Strumia,
Nucl. Phys. {\bf B674}, 171 (2003), hep-th/0305184.
%%CITATION = HEP-TH 0305184;%%

\bibitem{Buchbinder:2003qu}
I.~L. Buchbinder {\em et~al.},
Phys. Rev. {\bf D70}, 025008 (2004), hep-th/0305169.
%%CITATION = HEP-TH 0305169;%%

\bibitem{Luty:1999cz}
M.~A. Luty and R.~Sundrum,
Phys. Rev. {\bf D62}, 035008 (2000), 
hep-th/9910202.
%%CITATION = HEP-TH 9910202;%%

\bibitem{Gregoire:2004nn}
T.~Gregoire, R.~Rattazzi, C.~A.~Scrucca, A.~Strumia and E.~Trincherini,
%``Gravitational quantum corrections in warped supersymmetric brane worlds,''
Nucl. Phys. {\bf B720}, 3 (2005),
hep-th/0411216.
%%CITATION = HEP-TH 0411216;%%

\bibitem{Scrucca:2004cw}
C.~A.~Scrucca,
%``Mediation of supersymmetry breaking in extra dimensions,''
Mod.\ Phys.\ Lett.\ A {\bf 20} (2005) 297,
hep-th/0412237.
%%CITATION = HEP-TH 0412237;%%

\bibitem{Anisimov}
A.~Anisimov, M.~Dine, M.~Graesser and S.~Thomas,
%``Brane world SUSY breaking,''
Phys.\ Rev.\ D {\bf 65} (2002) 105011, 
hep-th/0111235;
%%CITATION = HEP-TH 0111235;%%
%``Brane world SUSY breaking from string/M theory,''
JHEP {\bf 0203} (2002) 036,
hep-th/0201256.
%%CITATION = HEP-TH 0201256;%%

\bibitem{Falkowski:2005zv}
A.~Falkowski, H.~M.~Lee and C.~Ludeling,
%``Gravity mediated supersymmetry breaking in six dimensions,''
hep-th/0504091.
%%CITATION = HEP-TH 0504091;%%

\bibitem{Gherghetta:2001sa}
T.~Gherghetta and A.~Riotto,
Nucl. Phys. {\bf B623}, 97 (2002), hep-th/0110022.
%%CITATION = HEP-TH 0110022;%%

\bibitem{Falkowski:2005fm}
A.~Falkowski,
%``On the one-loop Kaehler potential in five-dimensional brane-world
%supergravity,''
hep-th/0502072.
%%CITATION = HEP-TH 0502072;%%

\bibitem{wessbagger}
J.~Wess and J.~Bagger, ``Supersymmetry and supergravity'',
Princeton University Press (1992), Princeton, USA.

\bibitem{Buccella:1982nx}
F.~Buccella, J.~P.~Derendinger, S.~Ferrara and C.~A.~Savoy,
%``Patterns Of Symmetry Breaking In Supersymmetric Gauge Theories,''
Phys.\ Lett.\ B {\bf 115} (1982) 375.
%%CITATION = PHLTA,B115,375;%%

\bibitem{Procesi:1985hr}
C.~Procesi and G.~W.~Schwarz,
%``The Geometry Of Orbit Spaces And Gauge Symmetry Breaking In Supersymmetric
%Gauge Theories,''
Phys.\ Lett.\ B {\bf 161} (1985) 117.
%%CITATION = PHLTA,B161,117;%%

\bibitem{Gatto:1985jz}
R.~Gatto and G.~Sartori,
%``Zeros Of The D Term And Complexification Of The Gauge Group In
%Supersymmetric Theories,''
Phys.\ Lett.\ B {\bf 157} (1985) 389;
%%CITATION = PHLTA,B157,389;%%
%``Consequences Of The Complex Character Of The Internal Symmetry In
%Supersymmetric Theories,''
Commun.\ Math.\ Phys.\  {\bf 109} (1987) 327.
%%CITATION = CMPHA,109,327;%%
 
\bibitem{Luty:1995sd}
M.~A.~Luty and W.~I.~Taylor,
%``Varieties of vacua in classical supersymmetric gauge theories,''
Phys.\ Rev.\ D {\bf 53}, 3399 (1996),
hep-th/9506098.
%%CITATION = HEP-TH 9506098;%%

\bibitem{Dine:1995ag}
M.~Dine, A.~E.~Nelson, Y.~Nir and Y.~Shirman,
%``New tools for low-energy dynamical supersymmetry breaking,''
Phys.\ Rev.\ D {\bf 53}, 2658 (1996), hep-ph/9507378.
%%CITATION = HEP-PH 9507378;%%

\bibitem{Carpenter:2005tz}
L.~Carpenter, P.~J.~Fox and D.~E.~Kaplan,
%``The NMSSM, anomaly mediation and a Dirac Bino,''
hep-ph/0503093.
%%CITATION = HEP-PH 0503093;%%

\bibitem{Affleck:1983vc}
I.~Affleck, M.~Dine and N.~Seiberg,
%``Dynamical Supersymmetry Breaking In Chiral Theories,''
Phys.\ Lett.\ B {\bf 137}, 187 (1984).
%%CITATION = PHLTA,B137,187;%%
  
\bibitem{Meurice:1984ai}
Y.~Meurice and G.~Veneziano,
%``Susy Vacua Versus Chiral Fermions,''
Phys.\ Lett.\ B {\bf 141}, 69 (1984).
%%CITATION = PHLTA,B141,69;%%

\bibitem{Buchbinder:1998qv}
I.~L.~Buchbinder and S.~M.~Kuzenko,
``Ideas and methods of supersymmetry and supergravity: Or a walk through
superspace'', IOP (1998), Bristol, UK.
%\href{http://www.slac.stanford.edu/spires/find/hep/www?irn=4231406}{SPIRES entry}

\bibitem{Gates:1983nr}
S.~J.~Gates, M.~T.~Grisaru, M.~Rocek and W.~Siegel,
%``Superspace, Or One Thousand And One Lessons In Supersymmetry,''
Front.\ Phys.\  {\bf 58}, 1 (1983), 
hep-th/0108200.
%%CITATION = HEP-TH 0108200;%%

\bibitem{Linch:2002wg}
W.~D.~Linch, M.~A.~Luty and J.~Phillips,
%``Five dimensional supergravity in N = 1 superspace,''
Phys.\ Rev.\ D {\bf 68}, 025008 (2003),
hep-th/0209060.
%%CITATION = HEP-TH 0209060;%%

\bibitem{Agashe:2005vg}
K.~Agashe, R.~Contino and R.~Sundrum,
%``Top compositeness and precision unification,''
hep-ph/0502222.
%%CITATION = HEP-PH 0502222;%%

\bibitem{Hisano:1995cp}
J.~Hisano, T.~Moroi, K.~Tobe and M.~Yamaguchi,
%``Lepton-Flavor Violation via Right-Handed Neutrino Yukawa Couplings in
%Supersymmetric Standard Model,''
Phys.\ Rev.\ D {\bf 53}, 2442 (1996), hep-ph/9510309.
%%CITATION = HEP-PH 9510309;%%

\bibitem{Rattazzi:1995ts}
R.~Rattazzi and U.~Sarid,
%``Disoriented Sleptons,''
Nucl.\ Phys.\ B {\bf 475}, 27 (1996), hep-ph/9512354.
%%CITATION = HEP-PH 9512354;%%

%\cite{Baldini:2004dj}
\bibitem{Baldini:2004dj}
A.~Baldini,
%``Status of the MEG experiment,''
AIP Conf.\ Proc.\  {\bf 721}, 289 (2004).
%%CITATION = APCPC,721,289;%%

\bibitem{Hall:1985dx}
L.~J.~Hall, V.~A.~Kostelecky and S.~Raby,
%``New Flavor Violations In Supergravity Models,''
Nucl.\ Phys.\ B {\bf 267}, 415 (1986).
%%CITATION = NUPHA,B267,415;%%

\bibitem{Barbieri:1995tw}
R.~Barbieri, L.~J.~Hall and A.~Strumia,
%``Violations of lepton flavor and CP in supersymmetric unified theories,''
Nucl.\ Phys.\ B {\bf 445}, 219 (1995),
hep-ph/9501334.
%%CITATION = HEP-PH 9501334;%%  

\bibitem{Borzumati:1986qx}
F.~Borzumati and A.~Masiero,
%``Large Muon And Electron Number Violations In Supergravity Theories,''
Phys.\ Rev.\ Lett.\  {\bf 57} (1986) 961.
%%CITATION = PRLTA,57,961;%%

\bibitem{Masiero:2004js}
A.~Masiero, S.~K.~Vempati and O.~Vives,
%``Massive neutrinos and flavour violation,''
New J.\ Phys.\  {\bf 6} (2004) 202,
hep-ph/0407325.
%%CITATION = HEP-PH 0407325;%%

\end{thebibliography}
\end{document}